# Fractal Systems of Central Places Based on Intermittency of Space-filling

Yanguang Chen

(Department of Geography, College of Urban and Environmental Sciences, Peking University, Beijing 100871, P.R.China. E-mail: chenyg@pku.edu.cn)

**Abstract**: The central place models are fundamentally important in theoretical geography and city planning theory. The texture and structure of central place networks have been demonstrated to be self-similar in both theoretical and empirical studies. However, the underlying rationale of central place fractals in the real world has not yet been revealed so far. This paper is devoted to illustrating the mechanisms by which the fractal patterns can be generated from central place systems. The structural dimension of the traditional central place models is $d=2$ indicating no intermittency in the spatial distribution of human settlements. This dimension value is inconsistent with empirical observations. Substituting the complete space filling with the incomplete space filling, we can obtain central place models with fractional dimension $D<d=2$ indicative of spatial intermittency. Thus the conventional central place models are converted into fractal central place models. If we further integrate the chance factors into the improved central place fractals, the theory will be able to well explain the real patterns of urban places. As empirical analyses, the US cities and towns are employed to verify the fractal-based models of central places.

## 1. Introduction

Central place theory seeks to explain the relative size and spacing of human settlements, including cities and towns, as a function of people's economic activities, especially shopping behavior. In fact, geographers have long recognized that the functions of cities and towns as market centers, traffic centers, or administrative centers result in a hierarchical system of settlements. A central place can be defined as a settlement at the center of a region, in which certain types of products and services are available to consumers (King and Golledge, 1978; Knox and Marston, 2007). In other words, the dominant function of a central place is to provide market and supply services for the region. The tendency for central places to be organized in hierarchical systems and network structure was first explored by Christaller (1933/1966), and his ideas led to central place theory, which was consolidated and developed by Lösch (1940/1954). From then on, this theory went gradually beyond geography and influenced many related fields, including economics, sociology, city planning theory, and even physical geography.

The theory on central places concerns with the way that human settlements evolve, and are spaced out and organized regularly. Three concepts are basic for us to understand this theory, that is, order, range, and threshold. The *order* of a central place is determined by the size of the region which is served and in turn reacts on it. A higher-order central place serves a larger region, while a lower-order central place serves a smaller region. Thus we need another concept, *range*, indicating the spacing of settlements. The range of a central place function denotes the maximum distance which consumers will normally travel to obtain a particular product or service. Both order and range can be related to the third concept, *threshold* of central place function, which suggests the minimum market size required to make the sale of particular product or service profitable (Knox and Marston, 2007). The three concepts can be mathematically measured with population size, spatial distance (service area), and number of central place, by which we can bring to light the latent scaling relations of central place systems.

The work of central place studies was involved with a series of concepts of complexity theory such as self-similarity, scaling laws, and self-organization. A central place system can be seen as both a hierarchy with cascade structure and a network with self-similar properties. A hierarchy and a network actually represent different sides of the same coin. Hierarchical structure is a very significant notion for us to understand urban fractals (Batty, 2006; Batty and Longley, 1994; Frankhauser, 1998). On the other hand, the hierarchies of central places bear an analogy with the network of rivers (Chen, 2009; Woldenberg and Berry, 1967), and river networks have been demonstrated to take on fractal nature (LaBarbera and Rosso, 1989; Rodriguez-Iturbe and Rinaldo, 2001; Tarboton *et al*, 1988). So, central place networks may be of self-similarity. Theoretically, the texture of central place network models can be interpreted with fractal geometry (Arlinghaus, 1985). Empirically, the central places of southern Germany do follow the scaling laws indicative of fractal structure (Chen and Zhou, 2006). The fractal texture of central place models have been expounded by Arlinghaus (1993) and Arlinghaus and Arlinghaus (1989). However, the generation mechanism of fractal structure of real central place systems is still an outstanding problem remaining to be resolved.

The central place theory is based on the postulates such as an unbounded isotropic plain with a homogeneous distribution of the purchasing power. The regular hexagonal formation has long been criticized on both description and explanation in geography because we cannot find this kind of patterns from the systems of settlements in the real world. However, the essence of this theory rests with its prediction on hierarchical scaling law and the average coordination number of cities around an urban place (six). Both population growth and human interaction activity follow the scaling laws (Rozenfeld *et al*, 2008; Rybski *et al*, 2009). The average coordination number was frequently demonstrated to be close to 6 (Niu, 1992; Haggett, 1969; Ye *et al*, 2000). More and more facts and empirical observations lend further support to the predictions from central place theory (this paper will present several evidences).

In fact, fractal central place theory may be one of the channels for us to comprehend the potential links between human systems and physical systems (e.g. rivers). However, the classical central place models reflect the intermittency-free systems of human settlements, which predict a Euclidean dimension for urban space in a region ($d$=2). This is in contradiction to the fractal patterns of settlements in the real world (Batty and Longley, 1994; Chen, 2008; Frankhauser, 1994). This paper is devoted to probing into intermittency of space-filling of central place systems. Intermittency is a significant concept to understand urban development, especially at large scale (Manrubia and Zanette, 1998; Zanette and Manrubia, 1997). The rest of the article is structured as follows. Section 2 gives the growing fractal model of central places by introducing the ideas of incomplete space-filling and intermittency, and section 3 provides empirical evidences by applying the fractal scaling relations to the systems of urban places in the United States. Related questions are discussed in section 4, and finally, the writing will be concluded by summarizing this study. The main novel points of this paper are as follows. First, the incomplete space-filling process and intermittent structure predicting fractal patterns are introduced into the classical central place models. Second, a new fractal model of cities, Koch snowflake model, is derived from the central place network. Third, the average nearest distance method is proposed to testify the central place scaling in the real world. Last but not least, the hierarchical scaling based on central place fractals can be employed to estimate the friction coefficient of distance decay.

## 2. Model modification

### 2.1 Complete space-filling and Euclidean plane

The basic central place systems can be divided into three types: the $k$=3 systems indicting a *marketing-optimizing* case, the $k$=4 systems suggesting a *traffic-optimizing* situation, and the $k$=7 systems indicative of an *administrative-optimizing* situation (Figure 1). A central place network is also a hierarchy consisting of 7 levels of settlements (L, P, G, B, K, A, M), including cities and towns. Each central place possesses 6 coordination locations for other central places. Suppose that the central place classes are numbered $m$=1, 2, …, $M$ in a top-town order for simplicity (generally $M$=7±2 in practice). This differs from the traditional models of central place hierarchies which are numbered in the bottom-up order. Three necessary measures of human settlement are symbolized as follows: $N_m$ and $P_m$ denote respectively the number and the average population size of the central places in the $m$th class, and correspondingly, $L_m$ denotes the average distance between adjacent central places of order $m$. We can examine and bring to light the scaling relations among the number $N_m$, size $P_m$, and distance $L_m$.

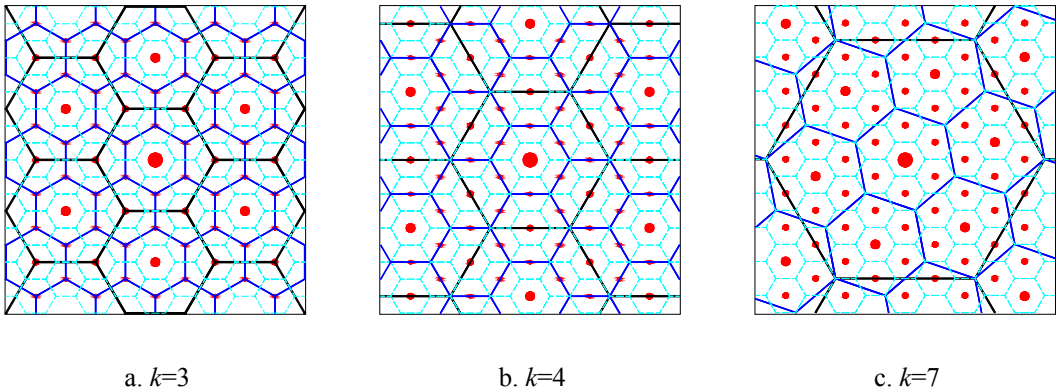

a. $k$=3  b. $k$=4  c. $k$=7

**Figure 1 Three intermittency-free central place networks representing different arrangements of human settlements**

The *nearest neighboring distance* $L_m$ should be explained as it is the first basic measure to define the fractal dimension of central place networks. In the $k=3$ networks, a central place of order $m$ have 3 central places of the same order as the nearest neighbors (Figure 1a). In the $k=4$ networks, for each central place in the $m$th level, the number of the most proximate central places of order $m$ is 4 (Figure 1b). In the $k=7$ networks, the number of immediate central places of each central place of order $m$ equals 5 (Figure 1c). The measure $L_m$ can be defined as the distance between a central place of order $m$ and the nearest neighboring central places in the same order.

Based on the number $N_m$ and distance $L_m$, the fractal dimension of central place networks can be defined mathematically (Chen and Zhou, 2004; Chen and Zhou, 2008). For simplicity, let $x=L_m$, $y=L_{m+1}$. For the $k=3$ network (Figure 1a), we have

$$(\frac{x}{2})^2 + (\frac{y}{2})^2 = y^2, \tag{1}$$

which is equivalent to

$$\frac{x}{2} = y\cos 30° = \frac{\sqrt{3}}{2}y. \tag{2}$$

Thus we get

$$x^2 = 3y^2, \tag{3}$$

which implies a scaling relation

$$x = \sqrt{3}y. \tag{4}$$

For $k=4$ network (Figure 1b), obviously we have

$$x = 2y = \sqrt{4}y. \tag{5}$$

As for $k=7$ network (Figure 1c), according to the cosine theorem, we get

$$x^2 = y^2 + (2y)^2 - 2y(2y)\cos 120° = 7y^2. \tag{6}$$

This suggests

$$x = \sqrt{7}y. \tag{7}$$

To sum up, a general scaling formula can be derived from the central place model (King and Golledge, 1978), and the result is

$$x = \sqrt{k}y. \tag{8}$$

So the *length ratio* of the average distance of order $m$ to that of order ($m+1$) is such as

$$r_l = \frac{x}{y} = \frac{L_m}{L_{m+1}} = \sqrt{k} \ . \tag{9}$$

The length ratio can also be called *distance ratio* in our context.

On the other hand, according to Christaller (1933/1966), the *number ratio* of the (*m*+1)th-order central places to that of the *m*th-order central places is as follows (Chen and Zhou, 2006)

$$r_n = \frac{N_{m+1}}{N_m} = k \ . \tag{10}$$

Then the similarity dimension of central place systems can be defined by the number ratio and the distance ratio in the form

$$D = -\frac{\ln(N_{m+1}/N_m)}{\ln(L_{m+1}/L_m)} = \frac{\ln(r_n)}{\ln(r_l)} = \frac{\ln k}{\ln \sqrt{k}} = 2 \ . \tag{11}$$

Equation (11) suggests the below spatial scaling relation such as

$$N_m = A L_m^{-d}, \tag{12}$$

where $A$ denotes the proportionality coefficient, and the subscript $m$ can be omitted for simplicity. The parameter $d=2$ refers to a Euclidean dimension, this implies that the geographical space is complete filling with human activities (Table 1, Figure 2). Therefore, the conventional central place models are based on a Euclidean plane despite the fact that the texture of the network comprises fractal lines (Arlinghaus, 1985).

Actually, Christaller's central place theory provided an *equilibrium solution* based on the thinking of classical economics to the problem that human settlements are organized and distributed spatially (King and Golledge, 1978). The network is regular and the distribution of the urban places in the same layer is homogeneous, thus the hexagonal market areas are the basic feature of central place distribution. As Prigogine and Stengers (1984, page 197) once observed: "Obviously, in actual case, such a regular hierarchical distribution is very infrequent: historical, political, and geographical factors abound, disrupting the spatial symmetry." The real systems of central places may be far from equilibrium (Allen, 1997). However, the network structure of regular hexagons may be broken down, but the spatial and hierarchical scaling relations hidden in the hierarchies are constantly held by the self-organized evolvement of settlements.

Table 1 Christaller's models for the hierarchies of central places in Southern Germany

| Type of place | Class $m$ | $k=3$ | | $k=4$ | | $k=7$ | |
|---|---|---|---|---|---|---|---|
| | | $L_m$ | $N_m$ | $L_m$ | $N_m$ | $L_m$ | $N_m$ |
| L | 1 | 108 | (1) | 256 | (1) | 1372 | (1) |
| P | 2 | 62.354 | 2 | 128 | 3 | 518.567 | 6 |
| G | 3 | 36 | 6 | 64 | 12 | 196 | 42 |
| B | 4 | 20.785 | 18 | 32 | 48 | 74.081 | 294 |
| K | 5 | 12 | 54 | 16 | 192 | 28 | 2058 |
| A | 6 | 6.928 | 162 | 8 | 768 | 10.583 | 14406 |
| M | 7 | 4 | 486 | 4 | 3072 | 4 | 100842 |

**Note**: The data come from Christaller (1966), see also King (1984). $N_m$ refers to the number of tributary areas, and $L_m$ to the distance between places (kms).

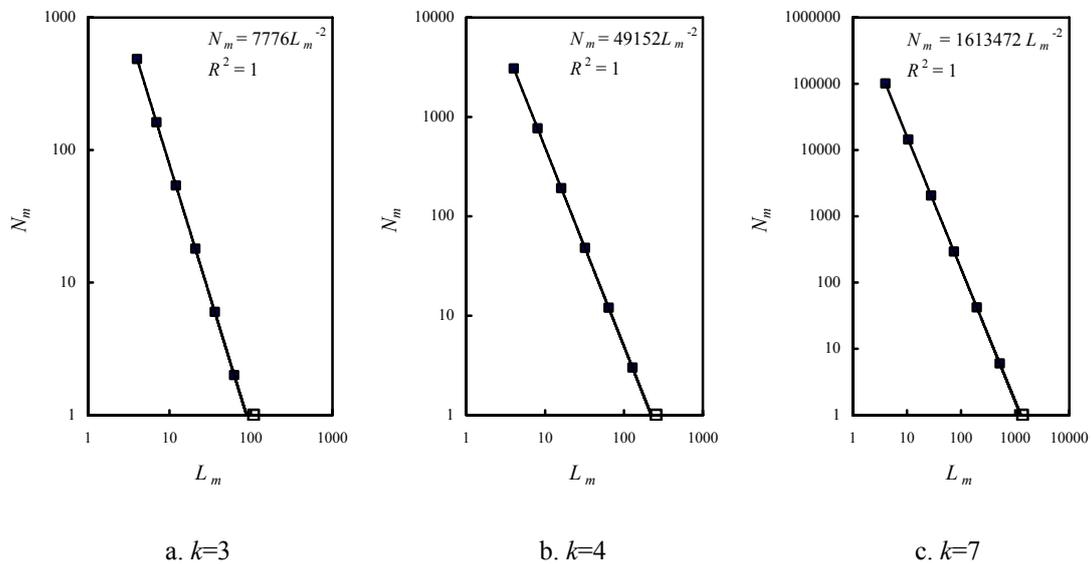

a. $k=3$     b. $k=4$     c. $k=7$

**Figure 2 The scaling relations between the number of tributary areas and the distance between places in the models of Southern Germany**

(**Note**: The first class is special so that it always goes beyond the scaling range. Therefore, the L type of central places is excluded from the scaling relations as outliers. See Chen and Zhou, 2006)

## 2.2 Incomplete space-filling and fractal systems

In the classical theory, one of the basic aims of central place systems is to make the best of geographical space. A network of settlements in a region can be grouped into several layers/levels in light of size and function of central places. Theoretically, the area served by a central place is a round field (Figure 3a). The areas served by different central places in the same layer are equal to one another. Because of competing with each other for trade/service area, round fields dominated by lots and lots of central places crowd and change to the close-packed hexagonal nets with cells of fixed diameter within a layer (Figure 1). The hexagonal network structure was assumed to avoid gaps or overlapping market area. In short, the traditional central place models indicate the intermittency-free systems of settlements making use of every bit of geographical space. There is no room for physical systems, and human activities are omnipresent in a territory.

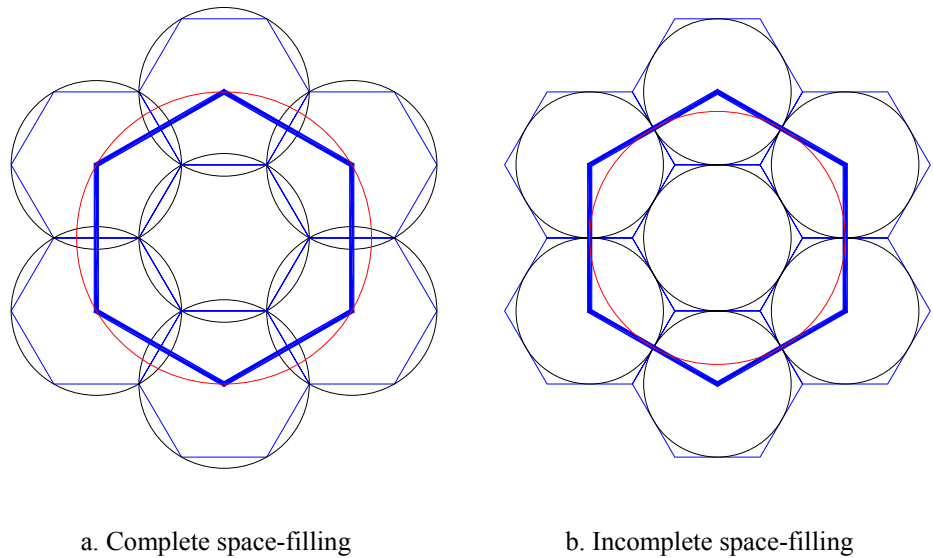

a. Complete space-filling        b. Incomplete space-filling

**Figure 3 The sketch maps of the complete space-filling and the incomplete space-filling ($k=4$)**
(**Note**: The hexagons represent the networks of central places, while the circles indicate the space-filling extent of human activities.)

However, this pattern of complete space-filling from spatial competition is only of ideality but never accord with the reality. The real systems of central places in Southern Germany, studied by Christaller (1933/1966) are in fact fractal systems with fractional dimension, and the fractal dimension values ranges from $D≈1.48$ to $D≈1.84$ (Chen and Zhou, 2006). In order to interpret this phenomenon, we need new assumption of incomplete space-filling associated with intermittency. Figure 3a refers to a scheme of complete space-filling process, while Figure 3b to a scheme of incomplete space-filling. The similarities and differences between the two schemes are very clear. The former suggests close-packed systems of service areas, while the latter suggests interspace representing the place which cannot or should not be occupied by human beings despite the compact distribution of urban places. Therefore, the revised model indicates an intermittent system of urban settlements with fractional dimension (see Appendix 1).

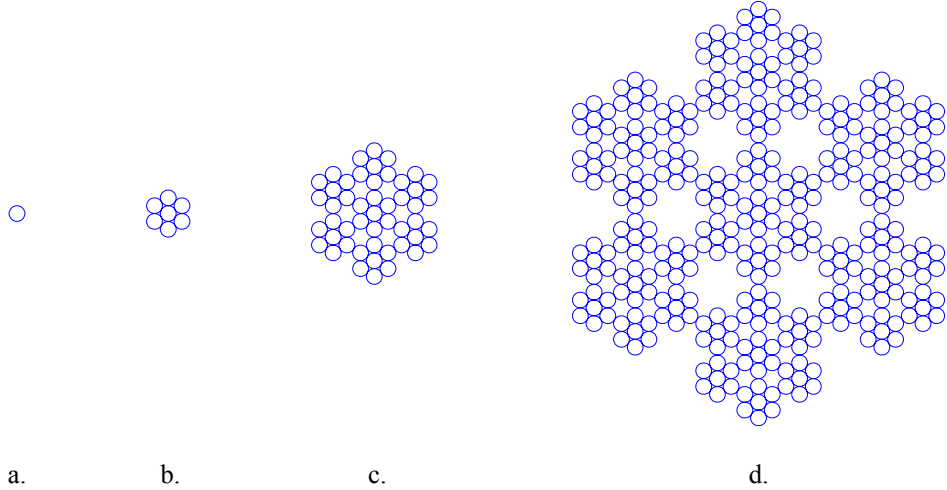

|  a. | b. | c. | d. |

**Figure 4 A growing fractal of central places based on incomplete space-filling and intermittency (the first four steps)**

Owing to the small but pivotal revision of the postulate on space-filling, the models of central places are improved. The central place networks can be transformed into a Koch snowflake pattern of systems of cities and towns, which is termed "urban snowflake" (Figure 5). From the standard central place models, we can naturally derive a Koch snowflake model (Chen, 2008; Chen and Zhou, 2006). The urban snowflake model (USM) is similar to a growing fractal presented by Jullien and Botet, (1987). The similarity dimension of the spatial form of the revised central place network is

$$D_n = \frac{\ln N_m}{\ln L_m} = \frac{\ln(N_{m+1}/N_m)}{\ln(L_{m+1}/L_m)} = \frac{\ln(7)}{\ln(3)} \approx 1.7712,$$

where $N_m$ refers to the number of central places of order $m$, and $L_m$ to the distance between two adjacent central places of the same order. The boundary of an urban fractal is possibly a fractal line, which contains useful geographical spatial information (Batty and Longley, 1987; Frankhauser, 1998). The perimeter of USM is a Koch curve with a dimension such as

$$D_b = \frac{\ln B_m}{\ln L_m} = \frac{\ln(B_{m+1}/B_m)}{\ln(L_{m+1}/L_m)} = \frac{\ln(4)}{\ln(3)} \approx 1.2619,$$

where $B_m$ denotes the boundary length of the urban snowflake of order $m$. The relation between the number of places and the boundary length of the urban snowflake follow the allometric scaling law

$$N_m \propto B_m^{\alpha} = B_m^{D_n/D_b}, \tag{13}$$

where $\alpha$ refers to the allometric scaling exponent, which is given by

$$\alpha = \frac{D_n}{D_b} = \frac{\ln N_m}{\ln B_m} = \frac{\ln(N_{m+1}/N_m)}{\ln(B_{m+1}/B_m)} = \frac{\ln(7)}{\ln(4)} \approx 1.4037.$$

Actually, there exist various allometric relations in the real systems of central places. Now, equation (12) can be generalized to the following form

$$N_m = AL_m^{-D_n}, \tag{14}$$

where $D_n$ indicates the fractal dimension of central place network. Equation (14) represents the spatial scaling law of central places. As soon as the intermittency is introduced to the central place systems, more than one kind of fractal structure can be derived from the common models by using the idea of fractals. No matter what kind of model it is, the urban place will follow the spatial scaling law.

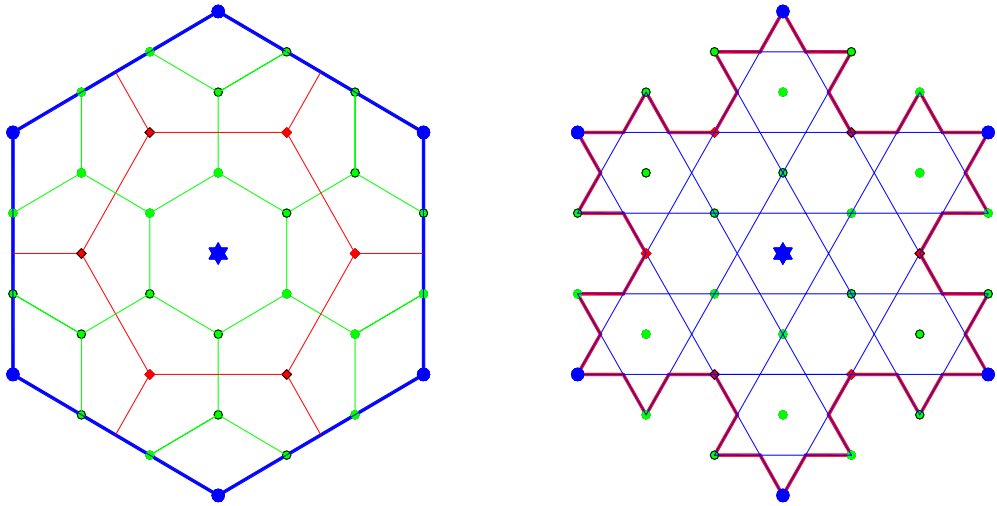

a. Central place network (CPN)　　　　　　b. Koch snowflake model (KSM)

**Figure 5 A sketch map of derivation of the Koch snowflake from central place network ($k=3$) (the first four levels, see Chen and Zhou, 2006)**

## 3. Empirical evidences

### 3.1 Cases of large scale

The central place models of Christaller (1933/1966) are abstracted from the real networks of human settlements. The fractal dimension of standard central place models can be divided into two types: one is *textural dimension* characterizing the boundary lines of service areas; the other is *structural dimension* characterizing spatial distribution of urban places. Where the models are concerned, the boundary lines of the traditional central place systems are fractal lines, and the dimensions are $D=\ln(4)/\ln(3)\approx1.2619$ for the $k=3$ system, $D=\ln(3)/\ln(2)\approx1.585$ for the $k=4$ system, and $D=\ln(9)/\ln(7)\approx1.1292$ for the $k=7$ system (Arlinghaus, 1985). All these can be regarded as textural dimension value. The structural dimension of traditional models is defined by equation (12), which suggests a Euclidean dimension $d=2$. However, the real networks of central places are of fractional dimension. Using the original data given by Christaller (1933) and the later supplemental data, Chen and Zhou (2006) estimated different fractal dimension values for central places in Southern Germany, that is, $D_n\approx1.7328$ for Munich region, $D_n\approx1.6853$ for Nuremberg region, $D_n\approx1.8370$ for Stuttgart region, and $D_n\approx1.4812$ for Frankfort region. The dimension of urban population distribution in Southern Germany in 1933 is about $D_p\approx1.8519$.

Now, we can apply the model of fractal central places to other countries such as America. In order to avoid the subjectivity of data processing, we can use the data processed by other scholars who know little about fractals. For making a multisided investigation, we should introduce the scaling relation between the number and average size of urban places of the same class. From the size distribution of the central place population formulated by Beckmann (1958), we can derived the size ratio of the urban population of order $m$ to that of order $m+1$, that is

$$r_p = \frac{P_m}{P_{m+1}} = \frac{s}{1-h}, \qquad (15)$$

where $s$ refers to the number of equivalent urban places of the ($m+1$)th level that are served by the $m$th level city, and $h$ to a proportionality factor that relates the urban population to the total population served by that city. Both $s$ and $h$ are assumed to be constant over the levels of the hierarchy. Please note the bottom-up order in the classical central place models is replaced by the top-down order in this paper for the purpose of simplicity of mathematical transformation, and this substitution does not influence the analytical conclusions.

A spatial and hierarchical scaling relations predicted by central place theory can be derived from equations (9), (10), and (15), and the results are as follows

$$P_m = CL_m^{D_p}, \qquad (16)$$

$$N_m = KP_m^{-\alpha}. \tag{17}$$

where $D_p$ denotes the fractal dimension of central place population, and $\alpha$ is the scaling exponent of the power-law relation between city number and average population size of urban places in the $m$th class. Both $C$ and $K$ are proportionality constants. Comparing equations (14) and (16) with (17) shows that

$$\alpha = \frac{D_n}{D_p}. \tag{18}$$

This suggests that $b$ is the ratio of the network dimension to the population dimension. In theory, the $\alpha$ value denotes the fractal dimension of city size distribution. It is equivalent in numerical value to the Pareto exponent and equals the reciprocal of Zipf's rank-size scaling exponent.

Now, let's examine the hierarchy of urban places in the United States. The data for the period 1900 to 1980 were processed by King (1984) who specialized in central places theory but seems to know little about fractals then (Table 2). Owing to absence of the distance measure, $L_m$, we can only probe the scaling relation between the city number and population size of urban places by fitting the data in Table 2 to equation (17). The results show that the US cities follow the scaling law on the whole. The last class indicating small urban places in 1960 and 1980 slop over the scaling range due to undergrowth of human settlements. As is often the case, the power-law relations break down when the scale are too large or too small (Bak, 1996). The urban centers with population size under 2000 can be treated as the so-called lame-duck class (Davis, 1978).

The lease squares calculations yield the four mathematical models taking on power-law relations. The equations, estimated parameter values, and related statistic quantities are listed in Table 3 for comparison. The effect of data points matching with the trend lines are displayed in Figure 6. From 1900 to 1980, the scaling exponent $\alpha=D_n/D_p$ varied around 1. This suggests that, at the large scale, the dimension of networks of urban centers is very close to that of the population distribution all over the central place systems. The same scaling analysis can be applied to Indian cities based on the census data from 1981 to 2001, and the effect is more satisfying (Appendix 2).

Table 2 The size scale and number of urban places in the United States, 1900-1980

| Class $m$ | Population size | Lower limit of size $P_m$ | Number of places $N_m$ | | | |
|---|---|---|---|---|---|---|
| | | | 1900 | 1940 | 1970 | 1980 |

| 1 | Over 1000000 | 1000000 | 3 | 5 | 6 | 6 |
| 2 | 500000--1000000 | 500000 | 3 | 9 | 20 | 16 |
| 3 | 250000--500000 | 250000 | 9 | 23 | 30 | 34 |
| 4 | 100000--250000 | 100000 | 23 | 55 | 100 | 117 |
| 5 | 50000--100000 | 50000 | 40 | 107 | 240 | 290 |
| 6 | 25000--50000 | 25000 | 82 | 213 | 520 | 675 |
| 7 | 10000--25000 | 10000 | 280 | 665 | 1385 | 1765 |
| 8 | 5000--10000 | 5000 | 465 | 965 | 1839 | 2181 |
| 9 | 2500--5000 | 2500 | 832 | 1422 | 2295 | 2665 |
| 10 | Under 2000 | 2000 | * | * | 627 | 1016 |

**Source**: The United States Bureau of the Census (1960, 1970, 1980). The Data is processed by King (1984). * Data unavailable. The last class, i.e., the 10$^{th}$ class, is a lame-duck class owing to undergrowth of cities.

**Table 3 The scaling models, scaling exponents, and corresponding goodness of fit for the US urban places, 1900-1980**

| Year | Mathematical model | Scaling exponent $b$ | Goodness of fit $R^2$ |
|---|---|---|---|
| 1900 | $N_m = 2347847.909 P_m^{-1.0053}$ | 1.0053 | 0.9909 |
| 1940 | $N_m = 4160392.9614 P_m^{-0.9813}$ | 0.9813 | 0.9931 |
| 1960 | $N_m = 12072505.9895 P_m^{-1.0243}$ | 1.0243 | 0.9786 |
| 1980 | $N_m = 22503822.0099 P_m^{-1.0708}$ | 1.0708 | 0.9736 |

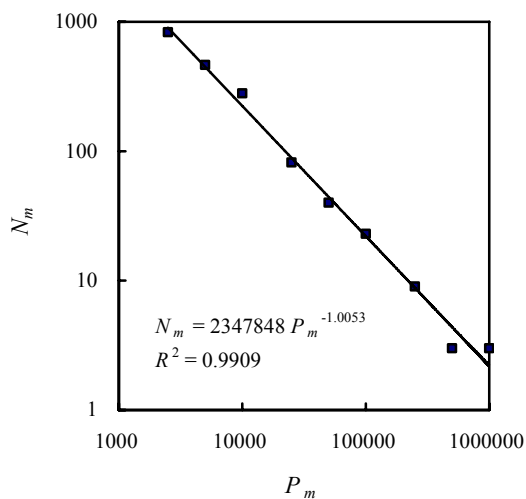
a. 1940

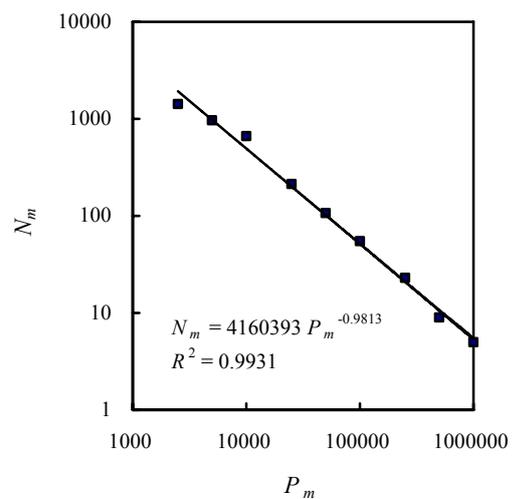
b. 1960

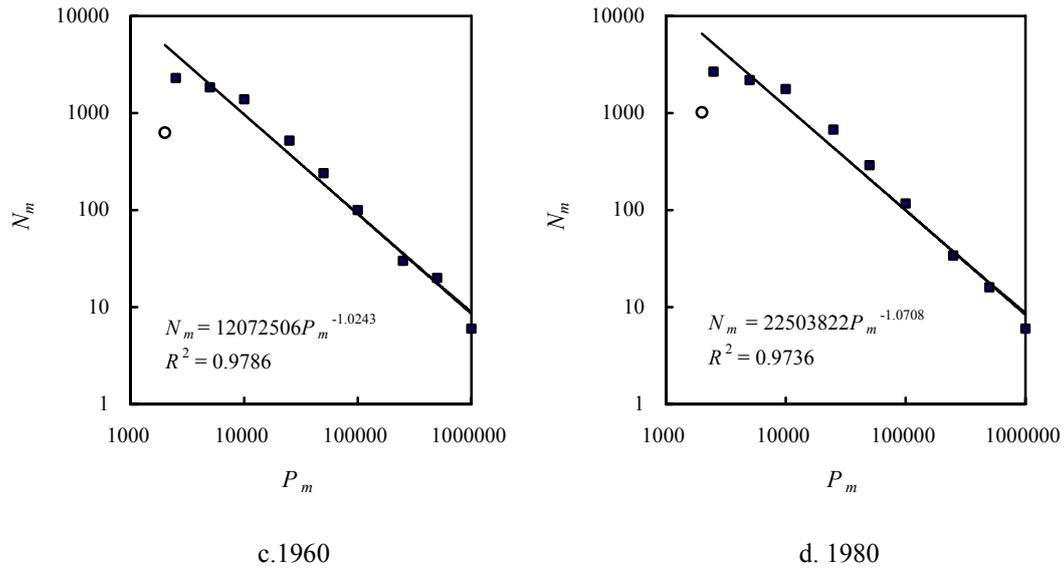

c. 1960    d. 1980

**Figure 6 The scaling relations between the population size and number of urban places in the United States, 1900-1980**

(**Note**: In 1960 and 1980, the last class is treated as an outlier because of undergrowth of cities. The circles indicate the exceptional data points out of the scaling ranges)

## 3.2 Cases of medium and small scales

More evidences can be found to support the scaling relation of fractal central places, equation (17), including the US cities in 2000 (Chen and Zhou, 2004), Chinese cities in 2000 (Chen and Zhou, 2008), India cities in 1981, 1991, and 2001(Basu and Bandyapadhyay, 2009), and the classical example of central places in southern Germany (Christaller, 1933). Next, let's turn to the urban places in the US sub-regions. Rayner *et al* (1971) once grouped the human settlements in the two US states, Iowa and North Dakota, into 7 or 9 classes in the bottom-up order in terms of the idea from central place theory. The distance refers to the mean distance to first nearest neighbor in miles. It is easy to rearrange the results in the top-down order according to our usage of classification (Table 4). The bottom level of the population size ranging from 10 to 99 is a lame-duck class due to undergrowth of urban places. These data can be employed to testify the fractal central place models.

**Table 4 The nearest-neighbor distance for different-sized urban places in Iowa and North Dakota**

| Class $m$ | Population size | Lower limit of size $P_m$ | Iowa | | North Dakota | |
|---|---|---|---|---|---|---|
| | | | Number $N_m$ | Distance $L_m$ | Number $N_m$ | Distance $L_m$ |

| 1 | Over 25000 | 25000 | 14 | 42.77 | 8 | 65.288 |
| 2 | 10000-24999 | 10000 | 11 | 30.71 | | |
| 3 | 5000-9999 | 5000 | 33 | 24.68 | 7 | 56.431 |
| 4 | 2500-4999 | 2500 | 46 | 22.51 | | |
| 5 | 1000-2499 | 1000 | 135 | 12.21 | 50 | 21.165 |
| 6 | 500-999 | 500 | 220 | 9.38 | 45 | 18.317 |
| 7 | 250-499 | 250 | 242 | 8.67 | 84 | 14.584 |
| 8 | 100-249 | 100 | 350 | 6.75 | 151 | 11.719 |
| 9 | 10-99 | 10 | 283 | 6.58 | 153 | 11.322 |

**Source**: Rayner JN, *et al*. 1971.

The least squares computations by fitting the data in Table 4 to equation (14), (16), and (17) yield two sets of results, which are listed in Table 5. The scaling relations are visually displayed in Figure 7. The fractal dimension of the network of human settlements, $D_n$, is about 1.8763 for Iowa and 1.7555 for North Dakota. These results are normal. However, the fractal dimension of the population distribution corresponding to these networks, $D_p$, are abnormal. The results are 2.7812 for Iowa and 2.2009 for North Dakota. The expected results are that the fractal dimension comes between 1 and 2. As similarity dimension, the values which are greater than the dimension of embedding space $d_E=2$ are understandable and acceptable (Chen, 2009). These suggest that the human settlements of small scale are concentrative, or the gap between the average distance of one class and that of its immediate class is not wide enough. The scaling exponent, $\alpha$, is supposed to be close to 1. But the value is less than what is expected because the population dimension $D_p$ exceeds the proper upper limit.

Anyway, urban fractals are evolutive processes rather than deterministic patterns. The fractal landscape of cities as systems and systems of cities often evolves from the nontypical into the typical form, and from the simple into the complex patterns (Benguigui et al, 2000; Chen, 2008). After all, human geographical systems differ from the classical physical systems. The models of human systems are dynamical modes instead of the static relations. Just because of this, we can optimize the spatial structure of human systems by using the ideas from fractals and related theories.

**Table 5 Mathematical models, scaling exponents, and goodness of fit for systems of urban places in Iowa and North Dakota (1971)**

| Region | Mathematical model | Scaling exponent $D$ or $b$ | Goodness of fit $R^2$ |
| --- | --- | --- | --- |
| Iowa | $N_m = 12708.1052 L_m^{-1.8763}$ | $D_n \approx 1.8763$ | 0.9564 |
|  | $P_m = 0.6768 L_m^{-2.7812}$ | $D_p \approx 2.7812$ | 0.9770 |
|  | $N_m = 11092.4287 P_m^{-0.6874}$ | $\alpha \approx 0.6874$ | 0.9397 |
| North Dakota | $N_m = 9878.1356 L_m^{-1.7555}$ | $D_n \approx 1.7555$ | 0.9799 |
|  | $P_m = 0.6904 L_m^{-2.2009}$ | $D_p \approx 2.2009$ | 0.9799 |
|  | $N_m = 3987.8406 P_m^{-0.7088}$ | $\alpha \approx 0.7088$ | 0.8707 |

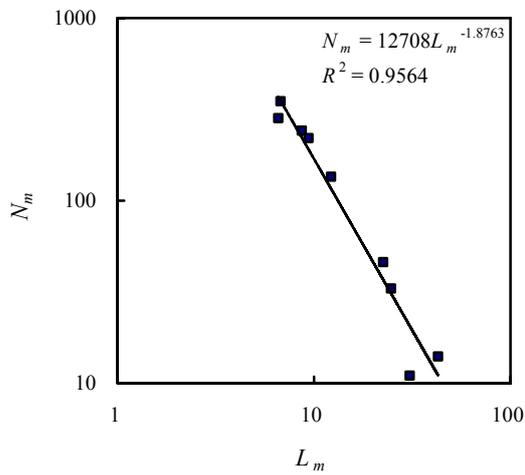

a. Iowa

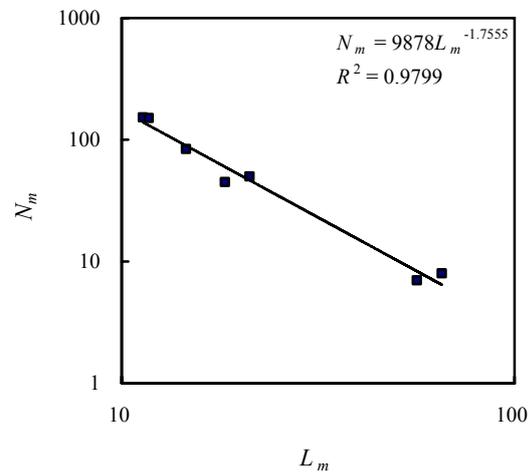

b. North Dakota

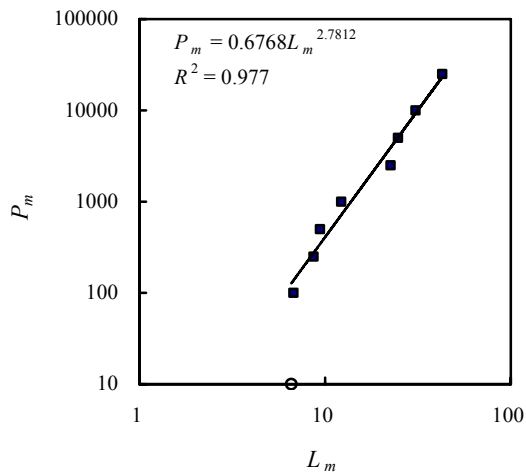

c. Iowa

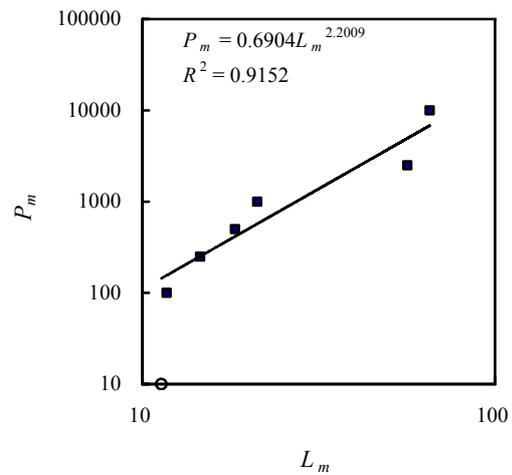

d. North Dakota

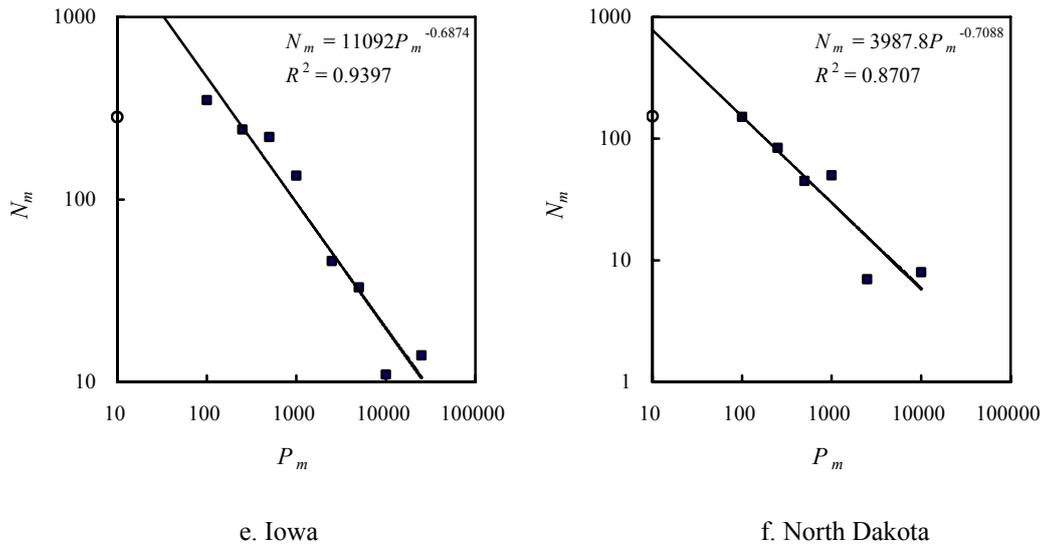

e. Iowa  f. North Dakota

**Figure 7 The scaling relations between numbers, population sizes, and mean distances to the first nearest neighbors of urban places, Iowa and North Dakota, USA (1971)**

(**Note**: In figures c, d, e, and f, the last class is an exceptional value owing to undergrowth of urban places. The circles indicate the outliers out of the scaling ranges)

## 4. Questions and Discussion

Central place theory is one of the cornerstones of human geography, and "our understanding of the growth and evolution of urban settlement systems largely rests upon the edifice of central place theory and its elaboration and empirical testing through spatial statistics" (Longley *et al*, 1991). Intuitively, the hierarchical structure of central place models is consistent with fractal geometry. However, the classical central place models cannot entirely explain the fractional dimension of systems of urban places in the real world. As soon as the idea of spatial intermittency is taken into account, the fractal structure of central place systems can be generated by recursive subdivision of space (Batty and Longley, 1994; Chen, 2008; Goodchild and Mark, 1987).

A key question is what is the dominative variable or parameter of the fractal dimension of central place systems. In order to answer this inquiry, let's make a simple mathematical transformation. Equation (9) can be generalized to the following expression

$$r_l = \frac{L_m}{L_{m+1}} = k^{1/w}, \tag{19}$$

where *w* is a positive number equal to or less than 2. Thus equation (11) changes to

$$D = \frac{\ln k}{\ln k^{1/w}} = w \leq 2. \tag{20}$$

This suggests that, in the simplest case, the scaling exponent of the ratio of the distance between the adjoining urban centers in one class to that in the immediate class controls the dimension of central place network. Further, suppose that equations (9) and (10) can be replaced with

$$r_l = \frac{L_m}{L_{m+1}} = k^{1/u}, \tag{21}$$

$$r_n = \frac{N_{m+1}}{N_m} = k^{1/v}, \tag{22}$$

where $u$ and $v$ are two parameters greater than zero. Thus the similarity dimension is such as

$$D = \frac{\ln r_n}{\ln r_l} = \frac{u}{v}. \tag{23}$$

This suggests that the dimension can be dominated by both the scaling exponent of numbers of urban places and that of distances between urban centers in different classes. In short, the parameters controlling the fractal dimension of networks are relative to the parameters of hierarchy of urban places. The structural dimension of central place systems is independent of the parameter $k$. This differs from the textural dimension, which depends on $k$ values of central place models (Arlinghaus, 1985).

Another question is why the average nearest-neighbor distance rather than the average distance is employed to estimate the fractal dimension of central place systems. As we know, to judge whether or not a geometric body is a fractal, we should drawn an analogy between the scaling exponent of the geometric body with Hausdorff's dimension. Hausdorff's dimension in pure mathematics can be replaced by box dimension in technique or technology. If we use the box-counting method to estimate the dimension of the geometric body, we should use the minimal/least box instead of the larger one to cover the body. Generally speaking, for dimension estimation, we should employ the boundary values rather the average value of scales to measure a geometric body. To estimate fractal dimension, we can adopt the upper limits of scales (the largest one) to make a measurement for the positive power law (PLR) relation or the area-radius scaling, or, the lower limits of scales (the smallest one) for the negative power law (NPL) relation or the box-counting scaling. If we substitute the scale mean for the scale boundary to make a measurement and thus to build a scaling relation, we may get a "fractal rabbit" (Kaye, 1989), rather than a valid fractal (see Appendix 3 for a simple example).

The hierarchical scaling laws derived from central place theory are very important for spatial analysis of cities. For example, we can use the scaling law to estimate the distance friction coefficient (DFC) of the urban gravity model (UGM). The principal parameter of UGM, actually a scaling exponent, is the DFC, which is hard to estimate in practice (Haggett *et al*, 1977). By means of the spatial and hierarchical scaling laws, we can derived a formula such as (the detailed derivation process will be given in a companion paper)

$$b = qD_n,  \qquad (24)$$

where $b$ refers to DFC, $q$ to the scaling exponent of Zipf's rank-size distribution, and $D_n$, the fractal dimension of spatial distance of cities defined above (Chen, 2008; Chen and Zhou, 2006). In terms of equation (18), we have

$$q = \frac{1}{\alpha} = \frac{D_p}{D_n}. \qquad (25)$$

Substituting equation (25) into equation (24) yields

$$b = \frac{D_n}{\alpha} = D_p. \qquad (26)$$

This suggests that if we measure the urban gravity with city population size, the average DFC value for the cities in a region is just the fractal dimension of spatial distribution of urban population. For example, this method can be applied to the urban places of Iowa and North Dakota discussed in Subsection 3.2. For the urban settlements in Iowa, the DFC value can be directly estimated as $d=D_p\approx2.7812$, or indirectly estimated as $d=D/b\approx1.8763/0.6874\approx2.7296$; for the urban settlements in North Dakota, the DFC value can be directly estimated as $d=D_p\approx2.2009$, or indirectly estimated as $d\approx1.7555/0.7088\approx2.4767$.

The third question is the type of central place fractals which is put forward preliminarily by Chen and Zhou (2006). If we consider the "shadow effect" of urban development, a variant of USM can be derived from the standard USM displayed in Figure 4. The shadow effect of regional science is proposed by Evans (1985, pages 98-99), who said: "In earlier years one would have expected hinterland effects to have dominated as the growth of the larger cities during the industrial revolution occurred at the expense of the smaller towns. In effect the large city cast it 'shadow' over the surrounding area depriving the smaller towns of growth as a larger tree prevents the growth of others by depriving them of light. But the extent of the 'shadow' and its effect will change as technical, social and economic factors change." Owing to the *shadow effect*, the small intermittency changes to large intermittency, and we have a hollow snowflake model (Figure 8a), which present a contrast to the solid snowflake models (Figure 8b). The network dimension ($D_n$), boundary dimension ($D_b$), and the allometric scaling exponent ($\alpha$) indicating the ratio of the two dimensions of the former are as follows

$$D_n = \frac{\ln(6)}{\ln(3)} \approx 1.6309, \quad D_b = \frac{\ln(4)}{\ln(3)} \approx 1.2619, \quad \alpha = \frac{\ln(6)}{\ln(4)} \approx 1.2925.$$

As a matter of fact, a majority of dimension values of central place networks in the real world vary from $D_n \approx 1.631$ for the hollow snowflake to $D_n \approx 1.771$ for the solid snowflake (Chen, 2008).

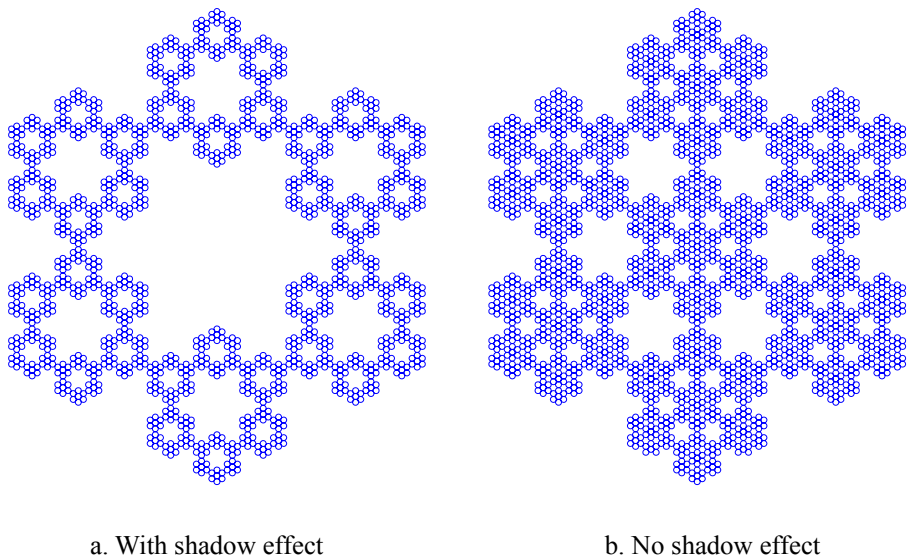

a. With shadow effect          b. No shadow effect

**Figure 8 Two kinds of urban snowflake models for central place networks with intermittency**

Anyway, central place theory is very important in theoretical geography, urban and regional economics, and city planning theory. One of the great triumphs of central place theory is that it implies the scaling law of spatial and hierarchical distribution of urban places, and the other, it predicts the city coordination number equal to six. What is more, it can be associated with fractals indicating spatial optimization. However, few systems of cities and towns in the real world can be fitted to the regular hexagonal patterns. Just because of this, central place theory has long been denounced on both institutional and methodological grounds. Fortunately, the problem used to puzzle geographers can nowadays be resolved to the core. First, by means of the hierarchical scaling law, we can improve central place theory so that it is based on fractal geometry rather than Euclidean geometry. Compared with Euclidean geometry, fractal geometry can go beyond graphics (figures) and yield more abstract mathematical equations. Second, if we introduce the chance factors into the regular central place fractals, we will obtain irregular central place fractal models, which look like the real settlement patterns. As a result, as done in Section 3, we can fit the observational data of urban settlements to the mathematical models rather than fitting the spatial distribution of urban places to the regular hexagonal network.

In order to develop central place theory, we should consider various factors affecting urban evolution such as chance, intermittency, and multi-scales. A fractal is a phenomenon of scaling symmetry consisting of form, chance, and dimension (Mandelbrot, 1983). Chance factor has been introduced to central place theory by Allen (1997), and intermittency factor is considered in this paper. As soon as both the chance for processes and intermittency for patterns are integrated into central place fractals, we will have brand-new central place models with irregular forms, the core of which is the spatial and hierarchical scaling and fractal dimension rather than hexagonal networks. The hexagonal structure can be treated as a postulate instead of the theoretical model itself. Thus the distance-based urban space concept will be replaced by the dimension-based urban space notion. In next step, USM should be generalized to multifractals since that central place hierarchies are associated with size distributions of human settlement and population distribution (Appleby, 1996; Chen and Zhou, 2006; Frankhauser, 2008), and the rank-size rule is related to multifractal phenomena (Chen and Zhou, 2004; Haag, 1994). As space is limited, the multifractal models of central place systems remain to be discussed in future studies.

## 5. Conclusions

Fractals suggest the optimized structure of systems in nature. A fractal body can occupy its space in the most efficient way (Chen, 2009; Rigon *et al*, 1998). Using ideas from fractals to plan cities and systems of cities will help to improving human environment and guaranteeing sustainable development of human society. Central place theory is one of the basic theories available for city planning. However, the theory is based on the concept of complete space-filling.

No buffer space, no vacant space, no intermittency for physical phenomena. In light of the traditional notion of economics, making the best of geographical space implies making use of geographical space in the best way. However, this notion may be old-fashioned. Going too far is as bad as not going far enough, and things will develop in the opposite direction when they become extreme. Complete space-filling suggests a simple Euclidean plane with little vital force and profound order. In contrast, fractal structure suggests the order behind chaos of cities (White and Engelen, 1993; White and Engelen, 1994). The unity of opposites of chaos and order may indicate driving force of urban evolvement.

The main points of this paper can be summarized as follows. First, the real systems of central places are of intermittency indicating incomplete space-filling of human activities. This is different from the classical models of central places based on complete space-filling and intermittency-free patterns. The central place systems are actually fractal systems taking on self-similar network and hierarchy with cascade process. Second, the regular hexagonal landscapes of the traditional models in ideality are broken down in reality, but the scaling relations behind presentational forms will keep and the spatial and hierarchical scaling laws dominate the spatio-temporal evolution of settlements. Third, the fractal landscapes of urban places are more acceptable for urban theory, resting with two aspects: one is that, by introducing the chance and intermittency factors into the hexagonal network, the fractal central place models accord with the real system of urban places and correspond to physical phenomena such as rivers, and the other is that fractals suggest optimal structure for human systems and provide potential application to planning systems of cities and towns in the future. Since fractal geometry can go beyond the limit of graphics, we can fit the observational data of urban settlements to the scaling laws, instead of fitting the real settlement pattern to the ideal hexagonal hierarchical systems, to test central place theory in the future.

## Acknowledgements

This research was sponsored by the National Natural Science Foundation of China (Grant No. 40771061).

order and chaos. *Chaos, Solitons & Fractals*, 1994, 4(4): 563-583

Woldenberg MJ, Berry BJL (1967). Rivers and central places: analogous systems? *Journal of Regional Science*, 7: 129-139

Ye D-N, Xu W-D, He W, Li Z (2001). Symmetry distribution of cities in China. *Science in China D: Earth Sciences*, 44(8): 716-725

Zanette D, Manrubia S (1997). Role of intermittency in urban development: a model of large-scale city formation. *Physical Review Letters*, 79(3): 523-526

## Appendices (The first and third appendixes can be deleted after review)

### A1 A sketch map of spatial intermittency

The spatial disaggregation is an important concept in theoretical geography. Spatial disaggreation includes recursive subdivision of space, hierarchy and network structure (Batty and Longley, 1994). The spatial subdivision can be divided into two types: intermittency-free subdivision (Figure A(a)) and intermittent subdivision (Figure A(b)). The former is associated with the Euclidean space while the latter with the fractal space. The classical central place models are based on the intermittency-free recursive subdivision of geographical space. This paper tries to introduce the idea of recursive subdivision of space with intermittency into central place theory and thus yield fractal central place systems.

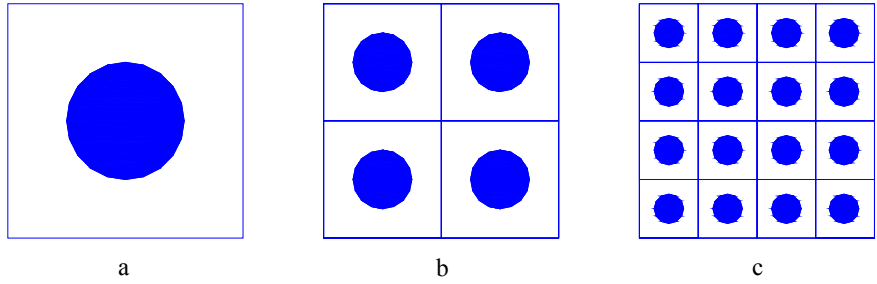

a. Spatial subdivision process without intermittency

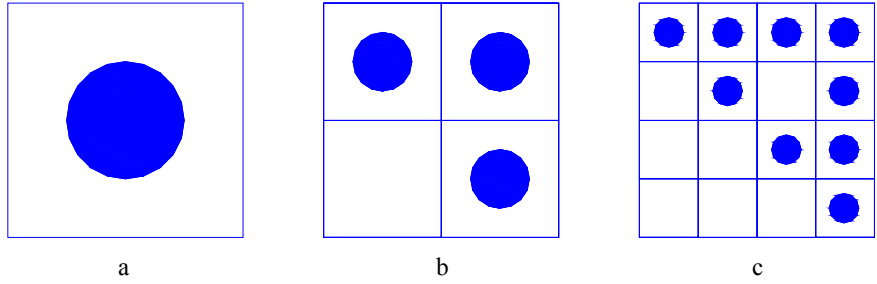

b. Spatial subdivision process with intermittency

**Figure A1 Sketch maps for two kinds of recursive subdivision of geographical space**

(**Note:** Figure A1 (a) displays a Euclidean process with dimension $d$=ln(4)/ln(2)=2; Figure A1 (b) shows a intermittent process with a fractal dimension $D_f$=ln(3)/ln(2)≈1.585)

## A2 Another case of large scale based on Indian cities

The same method of scaling analysis as that in Subsection 3.1 can be easily applied to Indian censuses of 1981, 1991 and 2001. The results are very satisfying (Tables A1 and A2, Figure A2). The case of India lends further support to the hierarchical scaling relation between city number and size predicted by central place theory.

**Table A1 The population size and number of Indian urban places: 1981-2001**

| Order | Population size | City number ($N_m$) | | |
| --- | --- | --- | --- | --- |
| ($m$) | ($P_m$) | 1981 | 1991 | 2001 |
| 1 | 50000 | 486 | 637 | 688 |
| 2 | 100000 | 216 | 296 | 427 |
| 3 | 200000 | 104 | 134 | 174 |
| 4 | 400000 | 49 | 73 | 91 |
| 5 | 800000 | 17 | 32 | 48 |
| 6 | 1600000 | 9 | 11 | 14 |
| 7 | 3200000 | 4 | 7 | 8 |
| 8 | 6400000 | 2 | 3 | 4 |
| 9 | 12800000 | * | * | 3 |

Source: The Indian censuses of 1981, 1991 and 2001. Cited from: Basu and Bandyapadhyay, 2009. * Data unavailable. The last class, i.e., the 9th class, is a lame-duck class owing to undergrowth of cities.

**Table A2 The scaling models, scaling exponents, and corresponding goodness of fit for Indian urban places, 1981-2001**

| Year | Mathematical model | Scaling exponent $b$ | Goodness of fit $R^2$ |
|---|---|---|---|
| 1981 | $N_m = 3073.3413 P_m^{-1.1472}$ | 1.1472 | 0.9982 |
| 1991 | $N_m = 3845.2462 P_m^{-1.1087}$ | 1.1087 | 0.9967 |
| 2001 | $N_m = 4093.4247 P_m^{-1.0528}$ | 1.0528 | 0.9900 |

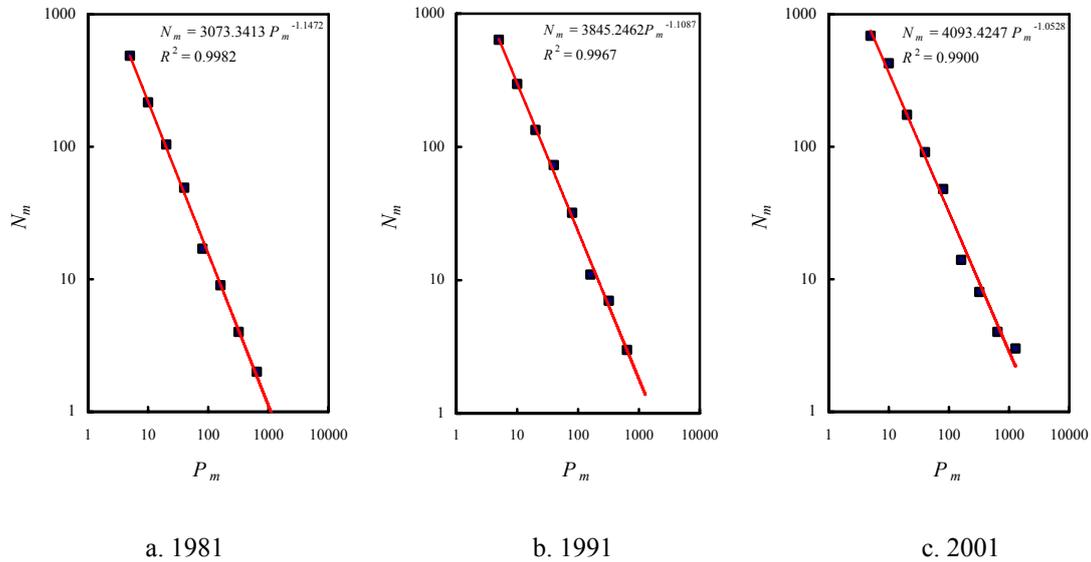

a. 1981      b. 1991      c. 2001

**Figure A2 The scaling relations between the lower limit of population size and number of urban places in India, 1981-2001**

### A3 Why to use the least average distance instead of the average distance

In order to calculate the dimension of a geometrical body, we should employ the boundary value instead of the mean value of scale to make a measurement. Concretely speaking, we use the upper limits of scales for positive power law (PLR) relation or the area-radius scaling to avoid improper scale translation, and use the lower limits of scales for negative power law (NPL) relation or the box-counting scaling to avoid box overlapping. The scaling exponent implies the dimension. For example, if we want to compute the dimension of a 2D plane, we can make use of the PLR-based area-radius scaling. Drawing a system of concentric circles yields a set of data (Figure A3, Table A3). Based on the upper limits of radius, a least squares computation gives the following result

$$A(r) = \pi r^d = 3.1416 r^2,$$

where $r$ refers to the upper limit of a radius for a ring, and $A(r)$ to the area of the corresponding ring, which is defined as the geometric region between two immediate circles. As for the parameters, $\pi$ denotes the ratio of the circumference of a circle to its diameter, and $d=2$ is just the Euclidean dimension of the geometric plane. This is a perfect fit ($R^2=1$). However, if we substitute the average radius for the upper limit of the radius, the result is as below

$$A(r) = 5.7751 r^{1.5935}.$$

This is a typical "fractal rabbit"! The goodness of fit is about $R^2=0.9928$, and scaling exponent is about 1.5935, not close to 2. Besides, the proportionality coefficient is not close to the *pi* value.

**Table A3 The scaling models, scaling exponents, and corresponding goodness of fit for Indian cities**

| Number | Lower limit of radius | Average radius | Ring area |
|---|---|---|---|
| 1 | 0.5 | 0.25 | 0.7854 |
| 2 | 1.0 | 0.75 | 3.1416 |
| 3 | 1.5 | 1.25 | 7.0686 |
| 4 | 2.0 | 1.75 | 12.5664 |
| 5 | 2.5 | 2.25 | 19.6350 |
| 6 | 3.0 | 2.75 | 28.2743 |
| 7 | 3.5 | 3.25 | 38.4845 |
| 8 | 4.0 | 3.75 | 50.2655 |
| 9 | 4.5 | 4.25 | 63.6173 |
| 10 | 5.0 | 4.75 | 78.5398 |

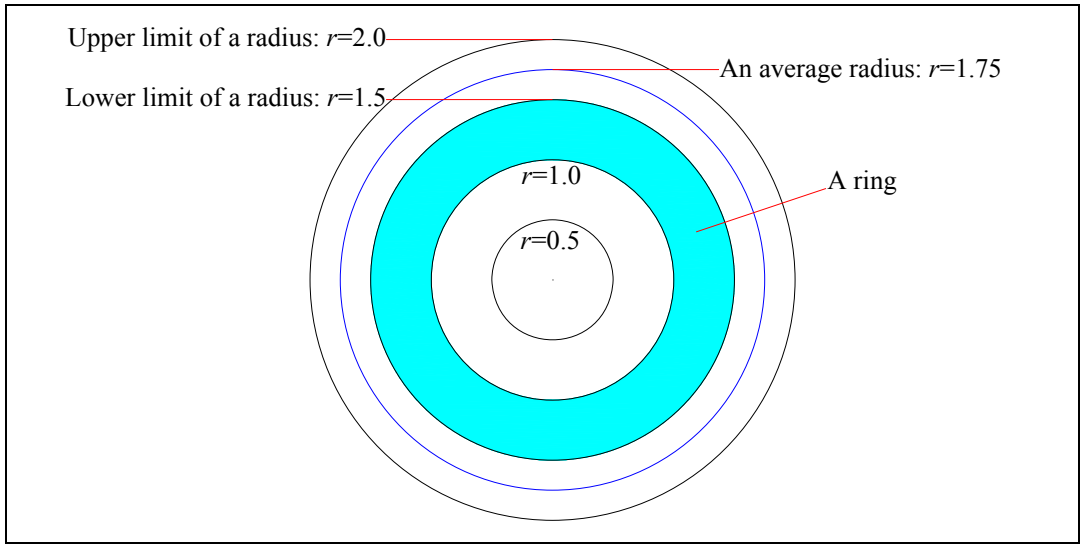

**Figure A3 A sketch map about the lower/upper limit of radii and average radii**